\documentstyle[12pt]{article}
\begin{document}
\begin{flushright}
TAUP-2298-95\\
hep-th/9510212
\end{flushright}

\baselineskip 24pt

\newcommand{\be}{\begin{equation}}
\newcommand{\ee}{\end{equation}}
\newcommand{\leqx}{\,\raisebox{-1.0ex}{$\stackrel{\textstyle <}
{\sim}$}\,}
\newcommand{\geqx}{\,\raisebox{-1.0ex}{$\stackrel{\textstyle >}
{\sim}$}\,}
\newcommand{\th}{\theta}
\newcommand{\va}{\varphi}
\newcommand{\de}{\Delta}
\begin{center}
{\bf Black Hole Information vs. Locality}
\\N.Itzhaki \footnote{Email Address:sanny@post.tau.ac.il}
\\Raymond and Beverly Sackler Faculty of  Exact Sciences
\\School
 of Physics and Astronomy
\\Tel Aviv University, Ramat Aviv, 69978, Israel
\end{center}
\begin{abstract}
We discuss the limitations on space time measurement in
the Schwarzchild metric.
We find that near the horizon the limitations on space time
measurement
are of the order of the black hole radius.
We suggest that  it indicates that a large mass black hole 
cannot be
described by means of local field theory even at macroscopic
distances and that any attempt to describe
black hole formation and evaporation by means of an effective
local field
theory  will necessarily lead to information loss.
We also present a new interpretation  of the black hole entropy
which leads to $S=cA$ , where $c$ is a constant of order $1$
which does
not depend on the number of fields.
\end{abstract}
\newpage
{\bf 1.Introduction}

The connection between black hole area and statistical mechanics
entropy is
 one of the most interesting question in physics.
Black hole analogs of the laws of thermodynamics exist
\cite{Hawking1,Hawking2} ,
 with the area of a classical black hole playing the role of
entropy.
 Furthermore , the generalized second law of thermodynamics
\cite{Bekenstein1,Bekenstein2,Bekenstein3} implies that the sum
of the ordinary
entropy and the black hole area (divided by 4 in  units where
$G=\hbar =c=1$)
 never decreases.
Finally, black holes radiate at the temperature ,
$T=\frac{dM}{dS}$ where
$S=\frac{A}{4}$ \cite{Hawking3,Hawking4}.
However, the statistical nature of the black hole entropy is
still unclear.
Another intricate related issue is the information loss puzzle
\cite{Hawking3}.
There seem to be three principal alternatives:

\hspace{-.85cm} 1. The information is lost \cite{Hawking3}.

\hspace{-.85cm} 2. The information is stored in the correlation
 between
Hawking radiation and Planck-mass remnants \cite{Aharon}.

\hspace{-.85cm} 3.All the information is emitted with  the Hawking
radiation \cite{'tHooft2,'tHooft3,'tHooft4}.

In this paper we study new aspects of the black hole
puzzle.
We consider a Gedanken experiment in which a macroscopic
system  falls into a black hole.
We are interested in the {\em absolute}  limitations on the
 information that an external observer can obtain on the system
{\em before} it reaches the horizon .
It has been argued by many authors
\cite{Mead,Sunny2,Maggiore,Veneziano,Dewitt,Unruh,Garay,Pad,Green}
 that Planck scale is the minimal observable scale.
In that case one should expect to find absolute limitations
 only after a time larger then $M\log (MR)$ (where $R$ is the
 initial location of the system) when the matter settles into a
 layer whose
invariant distance from the horizon is of the order of $1$.
So the radial information  is lost.
Note that  the angular information is not lost.
This problem is closely related to the trans-Planckian
 frequencies problem and to 't Hooft diagnosis that conventional
 quantum fields contribute  ultraviolet divergences to the entropy
 near the horizon; as such it cannot be investigated without
further
knowledge on quantum gravity
\footnote{Still one can learn on black holes from the fact that
 Planck
 scale is the minimal observable scale
since then  alternatives 1 and 2 are essentially the same
concerning the available information.
According to the remnants approach the information on the
collapsing star
 which form the black hole is in the correlation between the
 structure of
the remnant and Hawking radiation.
However, since the size of the remnants is of the order of
Planck scale one
cannot measure the structure of the remnants so the
information is
lost in principle.
In other words, although the state of the remnant and Hawking
 radiation
 is suppose to be a pure state.
One cannot distinguish between different states of the remnants.
One must, therefore, sum over all the possible states of the
remnants so
effectively one describes the state of Hawking radiation as a
mixed state.
We should note that this argument certainly does not support
previous
arguments \cite{'tHooft1,Susskind1} that remnants lead to
 mathematical inconsistency .}.
Recently, we have shown \cite{Sunny1} that there are states in
which the
uncertainties in space time measurements are much larger than
Planck scale.
This leads to  problems with the conventional definition of
statistical
 mechanics in quantum gravity even in the absence of  black holes.
In the presence of black holes this raises a new possibilities of
 information loss in   Gedanken experiments which can be studied
without further knowledge of quantum gravity.

{\bf 2.Space-time  measurements in Schwarzchild metric}

In this section we study the limitations on space-time
measurements in a
 Schwarzchild background metric , especially near the horizon.
First, we would like to consider a distant external observer who
 measures the
coordinate $(t, R, \th ,\va )$ of some space-time event.
In classical gravity the energy momentum tensor of the detector
which is
supposed to measure the location of the event can be as small as
 one
wishes and therefore it does not affect the metric.
So, as long as $R>2M$ there are no limitation in principle on
space time
measurements.
For an external observer, the time it takes for a  particle to
 reach the horizon
is infinite, thus at any finite time one can measure the exact
location and
momenta of the particles which fall into the black hole.
Thus in classical general relativity one can construct (at least
 in principle) a measuring device which follows the trajectory
 of the
particles in phase space even when a black hole is present, so
 information is not lost in principle.
The meaning of the no hair theorem in that context is that for
 practical
purposes the in falling particles are suddenly cut off from
communication
with the external observer, since the red-shift grows
exponentially in
 time near the horizon.
The only information which is available in practice is the total
 mass,
 angular momentum and charge while in principle all the
information is
available.

In quantum gravity the problem is much more interesting.
In the absence of a quantum theory of gravity, we need to
introduce two
postulates on quantum gravity in order to make the discussion
possible.

\hspace{-.85cm} The postulates are the following:

\hspace{-.85cm} {\em Postulate 1}. At large distances the first
order of the  gravitational
 effect in quantum gravity can be described to a good
approximation by
general relativity.

\hspace{-.85cm} {\em Postulate 2}. At large distances quantum
gravity is a local theory,
meaning there are no non-local effects at large distances.
We denote the minimal scale for which the postulates are correct
as
 $x_{c}$.
A few remarks are in order now.
First, since the only fundamental scale in quantum gravity is
$l_{p}$ ($ l_{p}=\sqrt{\frac{G\hbar}{c^3}}$) it is natural to
 assume that $x_{c}\approx l_{p}=1$.
Second, these postulates are the easiest way to describe the
correspondence
principle , between quantum gravity and general
relativity and between quantum gravity and local quantum field
theory.\footnote{Moreover, this correspondence principle is the
 only
experimental data that we have on quantum gravity.}
However, the postulates are not a general properties of quantum
 gravity
since there are more complicated way to describe the
correspondence
principle.
In section 4 we suggest that postulate 2 is incorrect and
that $x_{c}$ depends on the state of the system.
Third, in almost  any discussion on quantum gravity, one uses
those postulates.
In particular, the conventional argument for information loss
rests on
those postulates.

Let us focus first on the time measurement.
Basically we follow \cite{Sunny2} but in Schwarzchild metric
instead of
flat space-time.
In order to measure $t$ there must be a clock located at
$(R, \th , \va)$
which emits at least one photon towards the external observer.
There are two causes of error in this process of time measurement:

\hspace{-.85cm} 1- The uncertainty of the clock -$\de t$.

\hspace{-.85cm} 2- The uncertainty in the time it takes for the
photon to reach the
external observer . This uncertainty is due to the uncertainty
in the
metric caused by the uncertainty of the energy of the clock
$\de E$.

The time it takes for the photon to reach the external observer
 is
\be T=\int_{R}^{X} \frac{dr}{v}=(X-R)+2M_{tot}\log
(\frac{X-2M_{tot}}{R-2M_{tot}}),\ee
where $X$ is the location of the external observer and $M_{tot}$
is the
total mass of the black hole and the clock.
So we get
\be \de T>2\de E\frac{M_{b.h}}{\delta}\ee
where $\delta =R-2M_{b.h}$.
Adding the quantum uncertainty, the uncertainty for the whole
 process of time measuring is therefore
\be \de T_{tot}>\frac{1}{\de E}+2\de E
\frac{M_{b.h}}{\delta}\geq\sqrt{\frac{8M_{b.h}}{\delta}}.\ee
Note that the minimum is obtained for $\de
E\approx\sqrt{\frac{\delta}{M_{b.h}}}$, thus near the horizon
the clock is
rather regular (no planckian uncertainty).
This result is not surprising since the uncertainty in the
invariant
distance is of order one: $\de T_{tot}\sqrt{g_{0,0}}\approx 1$.
The limitation on $R$ measurement is
\be \de R_{tot}=\de T_{tot}\frac{\partial R}{\partial T}.\ee
{}From Eq.(1) we get $ \frac{\partial R}{\partial T}=
\frac{2M}{\delta}$ thus
\be \de R\geq\sqrt{\frac{\delta}{M}}.\ee
Again this is not surprising since the uncertainty in the
invariant
distance is of order one: $\de R\sqrt{g_{r,r}}\approx 1$.

There are at least two different ways to measure $\th $ or
$\va $ which
leads to the same surprising result.
The first is based on the logic of Eq.(4) but with $\va$ or
$\th$ instead of $R$.
The other way is more straight forward and is the following:
Suppose that there is an apparatus which send a light signal
from the event
in the radial direction.
Classically $\frac{d\th}{dt}=\frac{d\va}{dt}=0$ so $\th_{f}=\th$
 and
$\va_{f}=\va$.
The external observer need therefore to measure only $\th_{f}$
and $\va_{f}$ in
order to know $\th$ and $\va $.
Let us treat this process  quantum mechanically.
Consider the Schwarzchild metric
\be d\tau ^2=B(r)dt^2-A(r)dr^2-r^2(d\theta^2+sin^2 d\varphi ^2).
\ee
Where $B(r)=1-\frac{2M}{r}=A(r)^{-1}$.
According to postulate 1 we can use this metric at scale larger
 than $1$.
The  geodesic equations are \cite{Weinberg}
\be r^2\frac{d\varphi}{dp}=J\;\;\;\mbox{(constant)}\ee
\be A(r)(\frac{dr}{dp})^2 +\frac{J^2}{r^2}-\frac{1}{B(r)}=
-E\;\;\;\mbox{(constant)},\ee
where $p$ is a parameter describing the trajectory and $J$ is
the angular
 momentum per unit mass.
The connection between the proper time $\tau$ and $p$ is
\be d\tau ^2=E dp^2\ee
so we find that
\be E>0\;\;\;\mbox{for material particles}\ee
\be E=0\;\;\;\mbox{for photons}\ee
Since $A(r)$ is positive the particle can reach a radius $r$
only if
\be \frac{2J^2}{r^2}+E\leq \frac{1}{B(r)}\ee
which leads to
\be J^2\leq \frac{r^3}{r-2M}\equiv F(r)\ee
for photons. Let us find the upper limit on $J$ such that the
light signal  will reach an external observer.
The minimum of $F(r)$ is $27 M^2$ at  $\widehat{r} =3M$.
Thus, in order that  the photon will cross $\widehat{r}$ we must
impose
\be J^2\leq 27 M^2.\ee
Now, suppose that the photon has uncertainty $\Delta \varphi$ at
the
emission point.
{}From postulate 2 and the  uncertainty principle we get
\be \Delta P_{\varphi}\geq \frac{1}{\Delta \varphi R},\ee
where $P_{\varphi}$ is the momentum of the photon in the
$\varphi $
direction.
In Minkowski space
\be V_{\varphi}=R\frac{d\varphi}{dt}=\frac{P_{\varphi}}{P}.\ee
where $P_{\varphi} $ is the momentum at the $\varphi$ direction
 and $P$ is
the total momentum, so  in Schwarzchild metric we get
\be \frac{d\varphi}{dt}=\frac{P_{\varphi}}{RP}\sqrt{B(R)}.\ee

Since
\be J=\frac{R^2}{B(R)}\frac{d\varphi}{dt},\ee
we obtain
\be \Delta J =\frac{R \de P_{\va}}{ P}\sqrt{\frac{1}{B(R)}}.\ee
{}From Eqs.(7,8) one obtains \cite{Weinberg}
\be \va _{f}=\va \pm\int \frac{A(r)^{\frac{1}{2}}dr}{r^2(\frac{1}
{J^2 B(r)}-\frac{1}{r^2})^{\frac{1}{2}}}\ee

The connection between the momentum of the photon and its total
 energy
(including the gravitational energy of the interaction between the
photon and the black hole)  differs from the connection in
Minkowski
space by the red-shift factor.
 \be E_{\gamma}=P\sqrt{B(r)},\ee
 thus we get
\be\de\va\geq \frac{1}{\de P_{\va}R}+\frac{\de P_{\va}}
{E_{\gamma}}\ee
so,
\be \Delta\va_{min}\geq \frac{1}{\sqrt{M E_{\gamma}}}\ee
Finally,  Eq.(1) implies that the maximal energy of the photon
 is \be
E_{\gamma}\leq \delta\ee
other wise the  Schwarzchild radius of the black hole and the
clock before the emission of the photon is larger then $R$ .
so we to get
\be \Delta\va_{min}\geq \frac{1}{\rho}\ee
where $\rho $ is the invariant distance from the horizon:
\be  \rho =\int_{2M}^{R} ds=\int_{2M}^{R}\frac{dr}
{\sqrt{1-\frac{2M}{r}}}=
\sqrt{8M(R-2M)}\ee
Notice that the minimum is obtained at $\de J=\frac{M}{\rho}$,
so Eq.(14) does
not play an important role in our discussion (according to the
postulates
the discussion is meaningful only for $\rho>1$) and the small
perturbation
approximation of Eq.(20) is valid.

The minimal uncertainty in the area of a sphere with invariant
distance
from the horizon $\rho$ is ,
\be \Delta A\geq \frac{1}{\rho }A.\ee
Notice that for $\rho \approx 1$ we get $\de A=A$.
This suggests that for external observer, all the angular
information
 of the  particles involved in the formation of the black hole is
lost\footnote{Except, of course, from  the total angular
momentum which
can be measured by measuring the Kerr metric at large distance.
 Here we
 assume that the total angular is zero.}.
We should remark that for $\rho >M$ one should consider the
regular
limitations on space time measurement \cite{Sunny2} which are
then non
 negligible and lead to $\Delta A\approx 1$.

Next, we turn to the measurement of the distance between two
events which occur
near the horizon  of an infinitely massive black hole.
In that limit the resulting geometry outside the event horizon
 is described
 by the Rindler metric.
\be  d\tau ^2=dT^2-dZ^2-dX^i dX^i.\ee
In terms of Schwarzchild coordinates it is given by
\be  d\tau ^2=(\frac{dt}{4M})^2\rho ^2-\rho ^2 -dX^i dX^i.\ee
The  Minkowski and Schwarzchild coordinates are related by
\[Z=\rho \cosh(\frac{t}{4M})\]

\be T=\rho \sinh(\frac{t}{4M}).\ee
We did not use Rindler metric in the distant observer
discussion since Rindler metric is a good approximation to
Schwarzchild metric only for $\rho \ll M$, while the invariant
 distance
between the external observer and the horizon is at least of the
 order of
$M$.
In Minkowski space the minimal uncertainty in $T$ is of order 1
 \cite{Sunny2} ,
therefore
\be \de t\approx \frac{M}{\rho},\ee
which is in agreement with Eq.(3).
Suppose that the two events ($a$ and $b$) occur at the same
$X^i$ and that
one wishes to measure $\rho=\rho _{a}-\rho _{b}$.
The measurement can be carried out in the following way:
A clock at $\rho _a$ measures the time $t_{i}$ when a photon is
 sent
towards $\rho _b$.
At the other   object there is a mirror which reflects the
photon back to
the first object, where the clock measures the time $t_{f}$ when
 a photon
arrives.
It is easy to find that
\be t_{f}-t_{i}=4M\ln(\frac{\rho_{a}}{\rho _{b}})\ee
Since there is a minimal uncertainty in $t$ there is also a
minimal uncertainty
in $\rho$,
\be \de\rho =\de t\frac{d\rho}{dt}\approx 1\ee
Suppose  that the two events  occur at the same $\rho$ and that
one wishes to measure the transverse distance $X_t$, where 
$X_t^2=(X^{i}_{a}-X^{i}_{b})
(X^{i}_{a}-X^{i}_{b})$.
For $X_t\gg \rho$ the time it takes for a light signal to travel from $a$ to $b$
is \cite{Susskind2}
\be t=8M\ln(\frac{X_t}{\rho }),\ee
so
\be \de X_t\approx\frac{1}{\rho}X_t.\ee


{\bf 3.Information loss and black hole entropy}

In field theories  which do not involve gravitation there is
information
loss in practice -i.e. the number of orthogonal states with the
same
macroscopic properties can never decrease.
In principle, however
  information is conserved, since in principle one can always
construct a measuring device which measures the exact state of
the system.
Therefore, in field theories which do not involve gravitation
\be S_{mat}=S_{phy}.\ee
Where $S_{mat}$ is the log of the number of orthogonal states
with the same
macroscopic properties and $S_{phy}$ is the log of the maximal
number of
orthogonal states with the same macroscopic properties which can
be
distinguished by means of local measurements.
Note that if one permits a  non-local interaction between the
measuring
device and the system (such as a Von-Neumann
 interaction) then one can distinguish between all states and
there
 is no information loss even in the presence of gravitation.
However the interaction between the measuring
device and the system as any other interaction must be local.

If  postulates 1 and 2 are correct then in quantum gravity there
are
cases for which
\be S_{mat}>S_{phy}\ee
In \cite{Sunny1} we present such a case using the weak field
approximation.
We argued that the basic reason for information loss in quantum
gravity
is the locality postulate (postulate 2)  and not the
horizon ( in the weak field case there is obviously no horizon).
The horizon , however , causes a strong red-shift which makes the
information loss more obvious then in the weak field case.
Let us demonstrate the information loss by considering
 a simple example: two orthogonal states of a particle
 which falls into a black hole .

\be <r, \th, \va\mid \phi _{1}>=R(r)
\Theta_{1} (\th )\Psi (\va )\ee
\be <r, \th, \va\mid \phi _{2}>=R(r)\Theta_{2}(\th )\Psi (\va )\ee
where $\Theta_{1}(\th )$ yields $\th =\pm\de\th $
and  $\Theta_{2}(\th )$ yields $\th =\pi\pm\de\th $
and $\de\th \ll \pi $.
Suppose that at $t=0$ $r\approx 3M$, so that  $\mid \phi _{1}>$
 and $\mid \phi
_{2}>$ are distinguishable by means of local interactions.
After a finite time of the order of $2M\log M$, the invariant
 distance between
the particles and the horizon is of the order of one, then from
 Eq.(25) we
learn that $\de \th=\pi $, thus the external observer cannot
distinguish
 between the two orthogonal states by means of local 
interactions.

Consider now a macroscopic system which at $t=0$ is located  at
 a distance
$R$ $ (R>3M)$, so that $S_{mat}=S_{phy}$.
After a finite time of the order of $M\log RM$ the invariant
 distance
between the object and the black hole is of  order $1$ and the
 object is
in the volume $0\leq\rho\leq 1$, $0\leq \th\leq\pi$,
 $0\leq \va\leq2\pi$.
But the minimal uncertainties are $\de \rho\approx 1$,
$\de \th\approx \pi$ and $\de\va\approx 2\pi$.
So one cannot distinguish between any orthogonal states with
the same
macroscopic properties.
This agree with the classical no hair theorem.
The difference is that in quantum theory unlike classical theory
 the
 absolute information loss is a smooth function of time and the
total
information loss takes a finite time.
Since the black hole is made out of collapsing objects it
 implies that for black
holes
\be S_{phy}=0.\ee
It is not surprising, therefore, that  postulates 1 and 2 lead
 to a non-unitarity
description of the formation and evaporation of black holes
since as we
found out information is lost in principle in the formation of a
 black hole
within a finite time.
There have been many attempts to shown that $S_{mat}=S_{b.h}$ so
 Hawking
radiation is just a statistical mechanics radiation.
We find this idea disturbing since, if our arguments are
 correct, then
 one cannot distinguish in principle among the states which are
counted in
 $S_{mat}$, so black hole entropy cannot be defined as the log
 of the number of orthogonal states which can be distinguished
 in principle but cannot be distinguished in practice (the same
 macroscopic properties).

What is, then, the  Bekenstein-Hawking entropy?
Recall that the special property of gravitation which causes
 information
 loss is the fact that in general relativity , unlike in any
 other theory,
the fields define distances .
It is only natural , therefore, to define the locality loss
entropy ($S_{l.l}$)
 as the number of orthogonal states of the metric that cannot be
 distinguished in principle by means of local measurements.
If there had not been limitations on space time measurement then
 $S_{l.l}$ would have been zero, since any orthogonal states of
the metric define different distances so in principle they are
distinguishable in that case.
However,  there do exist  limitations on space time measurements.
Therefore, there are
orthogonal states of the metric which define different distances
 in such
way that the difference is smaller or equal than the limitation
 on space
 time measurement, so $S_{l.l}>0$.
We suggest  that Bekenstein-Hawking entropy is $S_{l.l}$.
Let us calculate $S_{l.l}$ in the case of a black hole.
Clearly , the main contribution to $S_{l.l}$ is near 
the horizon ($\rho\approx 1$) where the uncertainty is maximal
, $\de A\approx A$.
Thus we need to count the number of orthogonal states of the metric for which
\be 0\leq A( g_{\mu\nu})\leq A_{0},\ee
where $A_{0}=16\pi M^2$.
In the large $M$ limit the gravitational radiation states are
 plane
 waves, thus
\be h_{\mu\nu}(x, t)=\frac{e_{\mu\nu}}{\sqrt{\omega V}}\sum_{k}
(a_{k}e^{ikx-wt}+a_{k}^{\dagger}e^{-ikx+wt}).\ee
Where $g_{\mu\nu}=\eta_{\mu\nu}+h_{\mu\nu}$, $k_1=\frac{n_1}{M}$,
 $k_2=\frac{n_2}{M}$ , $k_3
=n_3$ and $w=\sqrt{k_1^2 +k_2^2 +k_3^2}$.
$V$ is the volume of a thin sphere with width of the order $1$.
According to the postulates we can consider only $k_i\leqx 1$
hence  the
number of modes is of the order of $A_0$ and $w\approx 1$ so we
get,
\be h_{\mu, \nu}(x, t)=\frac{e_{\mu, \nu}}{\sqrt{A_{0}}}
\sum_{k}(a_{k}e^{ikx-wt}+a_{k}^{\dagger}e^{-ikx+wt}).\ee
Now,
\be A( g_{\mu, \nu})=\int \sqrt{g} dx_{1} dx_{2}\approx\int
(1+\frac{1}{2}h-\frac{1}{8} h^2)\ee
Using Eq.(50) we obtain
\be A( g_{\mu, \nu})\approx A_0-
\sum_{k_{1}, k_{2}} N_{k_1, k_2}.\ee
Since the number of ways to distribute $N$ quanta among $M$
 state is $\frac{(N+M-1)!}{(N-1)!M!}$ and the number of mode is of the order
of $A_0$ we get 
\be  S_{l.l}=cA_0,\ee
Since the whole discussion is based on qualitative arguments
(the uncertainty principle, $x_{c}\approx 1$), one should not
 expect to
calculate $c$.
Nevertheless, it is clear that $c$ does not depend on the number
 of
fields.

{\bf 4.Locality and  causality }

A physical theory is a theory which predicts  the results of
measurements.
 The following postulate is, therefore, somewhat  natural.

\hspace{-.85cm} {\em Postulate 3} The limitations on
 measurements in a
 physical theory follow naturally from  the mathematical
description of the
 theory.

The inspiration for the postulate is of course quantum
mechanics, where the
commutation relation
$ [X,P]=i\hbar $
leads to the uncertainty relation
$ \de X\de P \geq \frac{\hbar}{2}$  ,
which was originally  found using measurement arguments.
The measurement arguments alone do not prove that the mathematical
description of a  particle as a point in phase space is incorrect.
However, even without considering other difficulties in the
classical
description, (black body radiation, photo-electric effect,
 stability of atoms, etc.)
 one can argue that it does not make sense to describe a
particle as a
point in phase space ($\de X=\de P =0$) and to claim an
uncertainty relation
$ \de X\de P \geq \frac{\hbar}{2}$  only when a measurement take
 place.
Note that postulate 3 excludes information loss in principle.

According to  postulate 3  the limitations on space time
 measurements in
quantum gravity
are due to the mathematical description  of quantum gravity.
Obviously , local quantum fields theory is not the proper
mathematical
description of quantum gravity in
that case.
Furthermore an unavoidable conclusion from Eq.(25, 35) and
 \cite{Sunny1} is that
the non-locality scale is not bounded.
In other words, the non-locality scale is a function of the
 state of the
system; in the case of a black hole near the horizon the
non-locality scale is of the order
of the size of the black hole radius!
Therefore, postulate 2 cannot coexist with  postulate 3.
If postulate 2 is incorrect then in quantum gravity the
non-locality scale
depends on the state of the system!
It might be very difficult to find the
right mathematical tool to describe such a theory but it is
  hopefully possible.
On the other hand if  postulate 3 is incorrect then the whole
description
of physics by means of mathematics is meaningless since in that
 case
 there is no dictionary which connects between the mathematical
 description of physics and physics \cite{Sunny1}.
Therefore, in our opinion  postulate 3 is more fundamental than
 postulate 1.
In this paper we do not present the mathematical description which
leads to an unbounded
non-locality scale , but discuss qualitatively the
possibility .
We should remark that since the non-locality scale depends on
 the state
 of the system and it is one of the scales which define the
state of the system it seems that finding the proper description
 means finding
quantum gravity.

{}From Eq.(25) we find that the scale of non-locality is given by
\be d_{r}=1 \ee
\be d_{\va\th}=\left\{ \begin{array}{ll}
                     1   & \rho >M\\
                     \frac{M}{\rho } & \rho <M
              \end{array}
      \right.
\ee
where $d_{r}$ and $d_{\va\th}$ are the invariant non-local
distance in
the radial  and angular direction respectively.
Consistency of such a non-local theory implies non-locality in
 the time direction also.
Otherwise the theory will suffer from acausall propagation at
 macroscopic scales.
The non-locality in the time direction is then the time it takes
 for a
light signal to travel from $(r, \va, \th)$ to
$(r+\de r, \va+\de\va, \th+\de\th)$.
In Rindler metric this gives
\be \de t=8M\sinh ^{-1}{\frac{M}{2\rho ^2}}\ee
In that case causality is preserve locally.
Still, it does not mean that there is no  causality
 violation since $d_{\va\th}$ depends on $\rho$, so there
 might be global violation of causality.
Postulate 3 yields that the description of gravitation by means
of local
fields , $g_{\mu\nu}$ at a scale smaller than $d$ is incorrect
and that the
correct description is such that the minimal length is given by
 Eq.(55, 56).
In particular a particle which falls into a black hole  is
 spread in the
angular direction
 (since $\th$ and $ \va$ them selves are spread)
 according to Eq.(25).
\footnote{This spreading effect  obviously does not occur in the
 context of
point like particle which is described by local field theory.
In the context of string theory a similar effect was found by
Susskind
\cite{Susskind2}.
This  is quite surprising since our calculations and
postulates are rather model independent.}
fields at
there is no
 is
This spreading effect can causes a global violation of
 causality.
Consider a signal emitted from point $a$
$(\rho =\rho _{1}, \th =0, \va =0)$
towards point $b$ $(\rho _{1}, 0, \va _{1})$ (see Figure~1).
\begin{figure}
\begin{picture}(300,120)(0,0)
\put(190,30){\circle{40}}
\put(190,53){\circle*{1.5}}
\put(193,53){c}
\put(190,63){\circle*{1.5}}
\put(193,63){a}
\put(157,30){\circle*{1.5}}
\put(154,20){b}
\put(167,30){\circle*{1.5}}
\put(164,20){d}
\put(230,30){\vector(-1,0){20}}
\put(234,30){horizon}
\end{picture}
\caption{Causality implies that the nonlocality effects are 
such that $t_{ab}-\Delta t \leq t_{ac}+t_{cd}+t_{db}$.}
\end{figure}
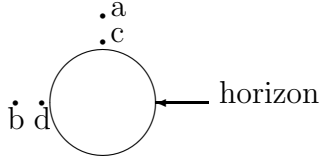
One can
calculate $t_{ab}$, the minimal time needed by  a classical
 perturbation to reach $b$ from $a$.
On the other hand due to the spreading effect one should consider
another path:
The signal is emitted from $a$ toward $c$ $(\rho_{0}, 0, 0)$
, ($\rho_0<\rho_1$) where it is spread with $\de\va \geq\frac{1}
{\rho_{0} }.$
Thus, if $\rho _{0}\leq \frac{1}{\va _{1}}$ then there is a
finite probability for the particle to be emitted from $d$
$(\rho_{0}, 0, \va _1)$ toward $b$.
Naively the condition for causality is
\be t_{acdb}\geq t_{ab}.\ee
However, we should also consider the non-locality in the $t$
direction,
meaning the uncertainty in $t$.
One can detect causality violation only if
\be t_{acdb}<t_{ab}-\de t_{ab},\ee
since according to postulate 3 time is defined with minimal error
 $\de t.$
In Rindler space it is easy to calculate $t_{ab}$ and $t_{ac}$,
\be t_{a, b}=8M\sinh ^{-1} (\frac{R}{2\rho_1})\ee
\be t_{ac}=t_{bd}=4M\log (\frac{\rho_{1}}{\rho_{2}})\ee
where $R$ is the distance between $a$ and $b$.
In order that path $acdb$ will be possible we must have
\be R\leq \frac{M}{\rho _{2}}\ee
thus
\be t_{ab}\leq 8M\sinh ^{-1} (\frac{M}{2\rho_1 \rho _{2}})\ee
Since $\de R=\frac{M}{\rho _1}$ we obtain
\be\de t_{ab}=8M\sinh ^{-1} (\frac{M}{2\rho _{1}^{2}}).\ee
So, the condition for causality is
\be 8M\log (\frac{\rho_{1}}{\rho_{2}})+8M\sinh ^{-1}
(\frac{M}{2\rho_1^2})\geq 8M\sinh ^{-1}
(\frac{M}{2\rho_1\rho_2}).\ee
Fortunately this condition is satisfied.
Equality is obtained
for $\rho_1 =M^a$ where $a<\frac{1}{2}$.
Further more if the spreading effect were stronger
 (for example $d=\frac{M}{\rho^c}$ where $c>1$) then causality
 would be violated.
The spreading rate (Eq.(55)) is , therefore, the fastest rate
consistent with causality.
\vspace{1.5cm}

I would like to thank  Prof. A. Casher and Prof. F. Englert
for helpful discussions.


\begin{thebibliography}{99}

\bibitem{Hawking1}S.W. Hawking, Comm. Math. Phys. 25 (1972) 152.
\bibitem{Hawking2} J.M. Bardeen, B. Carter and S.W. Hawking,
 Comm. Math.
Phys. 31 (1973) 161.
\bibitem{Bekenstein3} J. D. Bekenstein, Nuov. Cim. Lett. 4
(1972) 737.
\bibitem{Bekenstein1} J. D. Bekenstein, Phys. Rev. D 7, (1973)
 2333.
\bibitem{Bekenstein2} J. D. Bekenstein, Phys. Rev. D 9, (1974)
 3292.
\bibitem{Hawking3} S.W. Hawking, Phys. Rev. D14 ,(1976) 2460.
\bibitem{Hawking4} S.W. Hawking, Comm. Math. Phys. 43, (1975)
 199.
\bibitem{Aharon} Y. Aharonov, A .Casher and S. Nussinov, Phys.
Lett. B191 (1987) 51.
\bibitem{'tHooft2}C. R. Stephens, G. 't Hooft and B. F. Whiting,
 Class.
Quan. Grav. 11 (1994) 621.
\bibitem{'tHooft3} G. 't Hooft, ``Horizon operator approach to
black hole
quantization'', THU-94-02 gr-qc/9402037.
\bibitem{'tHooft4}G. 't Hooft, Nucl. Phys. B335 (1990) 138.

\bibitem{Mead} C.A.Mead Phys.Rev.135 B849 (1964).
\bibitem{Sunny2} N.Itzhaki  Phys.Lett.B 328 (1994) 274.
\bibitem{Maggiore} M.Maggiore Phys.Lett.B 304 (1993) 65.
\bibitem{Veneziano} D.Amati, M.Ciafaloni and G.Veneziano, Phys.
 Lett. B216
 (1989)41.
\bibitem{Dewitt} B.S.Dewitt, in Gravitation, ed. L.Witten (John
 Wiley \&sons
 New-york).
\bibitem{Unruh} W.G.Unruh, in Quantum Theory of Gravity, ed.
S.Christensen (Adam Hilger).
\bibitem{Garay} L.J. Garay, Int. J. Mod. Phys. A10 (1995) 145.
\bibitem{Pad} T. Padmanabhan, Gen. Relativ. Gravit. 17
 (1985) 215. 
\bibitem{Green} J. Greensite, Phys. Lett. B255 (1991) 375.
\bibitem{'tHooft1} G. 't Hooft, Nucl. Phys. B256, (1985) 727.
\bibitem{Susskind1} L.Susskind ``Trouble for remnants''
SU-ITP-95-1
hep-th/9501106.
\bibitem{Sunny1} N.Itzhaki ``Information loss in quantum gravity
 without
black holes'' TAUP-2250/95 hep-th/9508057 to appear in Class.
Quan. Grav.
\bibitem{Weinberg} S. Weinberg, Gravitation and Cosmology
 (John Wiley \& sons New-york 1972).
\bibitem{Susskind2} L.Susskind ``The world as a hologram''
 SU-ITP-94-33
hep-th/9409089.
``Black hole
\end{thebibliography}
\end{document}